\documentclass[aps,prd,amsmath,eqsecnum,showpacs,nofootinbib]{revtex4}
\usepackage{bm,amsfonts}
\usepackage{graphicx}
\newcommand{\rmd}{\mathrm{d}}
\newcommand{\rme}{\mathrm{e}}
\newcommand{\rmi}{\mathrm{i}}
\newcommand{\rmc}{\mathrm{c}}
\newcommand{\sgn}{\mathrm{sgn}}
\begin{document}
\title{Time evolution in quantum cosmology}
\author{Ian D. Lawrie}
\email{i.d.lawrie@leeds.ac.uk}
\affiliation{School of Physics and Astronomy, The University of Leeds, Leeds LS2 9JT, England.}
\date{\today}
\begin{abstract}
A commonly adopted relational account of time evolution in generally-covariant systems, and more
specifically in quantum cosmology, is argued to be unsatisfactory, insofar as it describes evolution
relative to observed readings of a clock that does not exist as a \textit{bona fide} observable object.
A modified strategy is proposed, in which evolution relative to the proper time that elapses
along the worldline of a specific observer can be described through the introduction of a `test clock',
regarded as internal to, and hence unobservable by, that observer.  This strategy is worked out in detail
in the case of a homogeneous cosmology, in the context of both a conventional Schr\"odinger quantization
scheme, and a `polymer' quantization scheme of the kind inspired by loop quantum gravity. Particular
attention is given to limitations placed on the observability of time evolution by the requirement that
a test clock should contribute only a negligible energy to the Hamiltonian constraint.  It is found that
suitable compromises are available, in which the clock energy is reasonably small, while Dirac observables
are reasonably sharply defined.
\end{abstract}
\maketitle
\section{Introduction\label{intro}}
Taken at face value, the canonical formulation of general relativity seems to entail that time evolution
is equivalent to a gauge transformation, and should therefore be physically unobservable.  Not surprisingly,
this `problem of time', along with other related interpretational issues, has attracted considerable
attention over many years.  The situation in the early 1990s was comprehensively documented in
\cite{ish,kuchar}; a recent survey is presented in \cite{anderson}, and textbook discussions can be
found in \cite{rovellibk,thiemannbk}. Everyday observations suggest that things actually do
change with time, and most investigators have concluded, in one way or another, that this fact can
be accomodated only by reference to physical clocks.  Many different schemes for implementing this general
idea have been proposed, but the notion of `relational', `emergent' or `internal' time, originating in the
work of Rovelli \cite{rovtime,rovref,rovpartial} has been quite widely adopted, especially in the context
of quantum cosmology, where many explicit calculations have become possible in recent years (see, e.g.
\cite{bojowald,ash09,ash10} for reviews). Roughly speaking, the relational-time approach involves identifying,
within the model considered, some quantity  that is to serve as a clock (in simple cosmological models, this is
typically a scalar field) and describing the evolution of other quantities relative to putative values
of this clock variable, which thus serves as an `internal time'. In this paper, we first argue that such accounts of
time evolution, while mathematically sound, are hard to interpret in a consistent manner, and cannot be the
whole story. We then propose a modified account, which we of course believe to be less unsatisfactory.

For concreteness, we discuss in section \ref{reltime} a particular implementation of the notion of internal
time given in \cite{aps1,aps2,acs} (an amended version of the last of these papers appears in \cite{acsrev},
and a rather different implementation, based on effective semiclassical dynamics is proposed in \cite{bojowaldetal})
and explain in some detail why we believe it to be deficient.\footnote{We emphasize that this discussion
is \textit{not} intended as a criticism of the work presented in \cite{aps1,aps2,acs,acsrev}.  On the contrary,
these papers describe an especially clear and thorough implementation of a view of time evolution that is now sufficiently
widespread to have something of an `official' status.  We do, of course, wish to suggest that
this `official' view is not wholly satisfactory.}   In brief, this is because (i) the physical
clock does not appear as an observable object in the final theory, so its putative readings cannot be
interpreted as values obtained by observation and have, indeed, no clear meaning; and (ii) no account is
available of the passage of time as experienced by real observers.

In sections \ref{conqu} and \ref{loopqu}, we describe, in the context of two different quantization
schemes, how these difficulties might be overcome, through the introduction of a `test clock'
associated with some specific observer (an idea proposed some time ago in \cite{le}). This makes it
possible to describe time evolution from the point of view of a particular observer, with respect to the
proper time that elapses along that observer's worldline.  A test clock should serve to reveal the
time evolution of the model universe while contributing negligibly to its energy content, and this
restricts the resolution with which the values of time-dependent observables can be determined. For both the
conventional `Schr\"odinger' quantization scheme of section \ref{conqu} and the loop-quantum-gravity-inspired
scheme of section \ref{loopqu}, we investigate this restriction, and find that suitable compromises are possible.
This proposal can be seen as a variant of the general idea of relational time, and we discuss its
relation to other versions in section \ref{discussion}.
\section{Relational time in quantum cosmology\label{reltime}}
The homogeneous, spatially flat cosmological model with no cosmological constant studied in
\cite{aps1,aps2,acs,acsrev} can be specified classically by the action
\begin{equation}
S=\int\rmd sN(s)^{-1}\left[-\frac{1}{24\pi G}\frac{(\partial_sv)^2}{v}+\frac{1}{2}v(\partial_s\phi)^2\right].
\label{action1}
\end{equation}
Here, $v(s)=a^3(s)$, where $a(s)$ is the usual Robertson-Walker scale factor, $N(s)$ is a positive, but
otherwise arbitrary lapse function and $s$ a correspondingly arbitrary time coordinate, while the matter
content is represented by the massless scalar field $\phi(s)$. We take this action to refer to a fiducial
cell of coordinate volume $\int \rmd^3x=1$, and $v$, with dimensions $(\textrm{length})^3$ to be the
physical volume of this cell. In SI units, $G=G_\mathrm{N}c^{-2}$, where $G_\mathrm{N}$ is the usual
Newton constant. The momenta conjugate to $v$ and $\phi$ are
\begin{eqnarray}
p_v&=&-\frac{1}{12\pi G}\frac{\dot{v}}{v}\\
p_\phi&=&v\dot{\phi},
\end{eqnarray}
where the overdot denotes differentiation with respect to the proper time
\begin{equation}
t(s)=\int_0^sN(s')\rmd s',\label{propertime}
\end{equation}
so $\dot{v}=N^{-1}\rmd v/\rmd s$, etc.
Variation of the action (\ref{action1}) with respect to $N$ yields the constraint $H_0=0$, where
\begin{equation}
H_0=-6\pi G vp_v^2+\frac{1}{2}v^{-1}p_\phi^2,\label{H0}
\end{equation}
is the generator of translations in $t$.

The problem of describing time evolution arises, as is well known, from the fact that \textit{bona fide}
gauge-invariant observables (Dirac observables) must commute with the constraint, and must therefore be
constants of the motion. The solution to this problem is often taken to involve identifying a variable
within the theory which can serve as a physical clock, and thus provide a notion of evolution with respect
to an `internal' or `emergent' time.  In \cite{aps1,aps2,acs,acsrev}, this relational picture is implemented
in the quantum theory by casting the constraint in the `deparametrized' form
\begin{equation}
\rmi\hbar\frac{\partial\Psi(\phi,v)}{\partial\phi}=-\sqrt{\hat\Theta(v,p_v)}\Psi(\phi,v),\label{evolv}
\end{equation}
where $\sqrt{\hat\Theta}$ is the square root of a suitably defined operator corresponding to
the classical expression $12\pi Gv^2p_v^2$.  The precise form of this operator depends on
the particular quantization scheme adopted.  This is formally similar
to a non-relativistic Schr\"odinger equation
\begin{equation}
\rmi\hbar\frac{\partial\Psi(t,x)}{\partial t}=\hat{H}(x,\partial_x)\Psi(t,x)\label{schrod}
\end{equation}
and appears to provide a notion of evolution with respect to an internal time represented by the scalar
field $\phi$.  (A much earlier study of quantum cosmology, also using a massless scalar field as an
internal time, was presented in \cite{blyth}.) In particular, an operator such as
\begin{equation}
\hat{V}(\phi_0):=\exp\left(-\rmi\sqrt{\hat\Theta}(\phi-\phi_0)/\hbar\right)
\hat{v}\exp\left(\rmi\sqrt{\hat\Theta}(\phi-\phi_0)/\hbar\right),\label{vofphi0}
\end{equation}
(where $\hat{v}$ acts by multiplication on $\Psi(\phi,v)$) is, for any fixed value of the parameter
$\phi_0$, a gauge-invariant Dirac observable:  if $\Psi(\phi,v)$ is a solution to the constraint
equation (\ref{evolv}), then $\hat{V}(\phi_0)\Psi(\phi,v)$ is another solution. In the language introduced
by Rovelli, $\hat{V}(\phi_0)$ is an `evolving constant of the motion' \cite{rovtime}, providing a
1-parameter family of `complete observables' \cite{rovpartial} labeled by $\phi_0$.  For the model
considered here, the classical solution for $\phi$ is always a monotonic function of $t$, so it is
tempting to interpret the constraint (\ref{evolv}) as effectively describing evolution with time,
in such a way that $\hat{V}(\phi_0)$ represents ``the volume at the time when the scalar field has
the value $\phi_0$''.  (Restrictions on the choice of variables that might serve as `internal time'
are discussed in \cite{gambiniporto}.) For the reasons we are about to present, we think that this
interpretation has significant limitations, and in subsequent sections we will suggest how
some of them might be overcome.

The limitations we have in mind are indicated by the following interrelated observations:
\begin{enumerate}
\item Despite their formal similarity, the constraint equation (\ref{evolv}) and the Schr\"odinger
equation (\ref{schrod}) do not mean the same thing.  In the case of a non-relativistic particle,
the wavefunction $\Psi(t,x)$ is, for each fixed value of $t$, an element of the physical Hilbert
space $\mathcal{H}_\mathrm{phys}=L^2(\mathbb{R},\rmd x)$ corresponding to a possible instantaneous
state of the particle.  A solution of (\ref{schrod}) yields a sequence of such states, labeled by the
external time parameter $t$, and in that apparently straightforward sense describes the time evolution
of the state of the 1-particle system.  By contrast, the physical Hilbert space of the cosmological
model is a space of \textit{solutions} of the constraint equation (\ref{evolv}); a solution of this equation
specifies not a sequence of possible states, but a single state characterized by a certain correlation
between `partial observables' $v$ and $\phi$.

\item Rovelli \cite{rovpartial} defines a `partial observable' as a quantity for which a measurement
procedure can be specified, in contrast to a `complete observable', whose value can be predicted by
theory.  He appears to take the view that the time parameter $t$ in (\ref{schrod}) is, in this sense,
a partial observable, but we disagree. The parameter $t$ does not refer to the reading of any physical
clock.  It is an external parameter, more like Newton's `absolute, true and mathematical time' or, as
described by Unruh and Wald \cite{unruh89,uw}, a `heraclitian time', which `sets the conditions' for
a measurement to be made. To be sure, the times recorded in a laboratory notebook during the course
of an experiment intended to test the validity of (\ref{schrod}) will refer to the readings of some
physical clock. But then, according to standard quantum mechanics, the state of the combined system
of a particle (position $x$) and clock (pointer reading $T$, say) is described by a wavefunction
$\Psi(t,x,T)$, governed by its own Schr\"odinger equation. Under suitable conditions, a sequence of
observed values of $T$ may closely approximate the corresponding values of $t$ at which the observations
were made, and one might derive an \textit{approximate} version of (\ref{schrod}) in which $t$ is replaced
with $T$.  However, this description is necessarily approximate, and there are well known restrictions
on the ability of  a quantum-mechanical clock to furnish a reliable measure of $t$ (see, e.g.
\cite{peres,hartle,uw,gambinipullin}).\footnote{More precisely, a wavefunction that realizes an exact
correlation between $T$ and $t$ must have the form $\Psi(t,x,T)=\delta(T-t)\psi(T,x)$; at each instant
$t$, it is an eigenfunction of the pointer operator $\hat{T}$ that acts by multiplication. The Schr\"odinger
equation admits solutions of this form only if the clock Hamiltonian is its conjugate momentum, $H_T=p_T$
\cite{hartle} which, being unbounded below, is physically unrealistic. Moreover, this wavefunction is
not in the physical Hilbert space $L^2(\mathbb{R}^2,\rmd x\rmd T)$, so even in this idealized case, the
possibility of identifying values of $t$ with the results of measurements made on the clock is doubtful.
No real clock will exist perpetually in a sequence of eigenstates of $\hat{T}$, so a Schr\"odinger equation of the form
$\rmi\partial_T\psi(T,x)=\hat{H}(x,\partial_x)\psi(T,x)$ gives at best an approximate, effective description
of the correlations exhibited by sequences of measured values $x$ and $T$, ignoring, for example, the loss
of unitarity resulting from the repeated measurements needed to obtain these values \cite{gambinipullin}.}
Consequently, if the scalar field $\phi$ in (\ref{evolv}) is taken as analogous to the reading $T$ of a
laboratory clock, we should expect that equation to be only approximately valid.  That is not so, however:
(\ref{evolv}) is an exact constraint equation, not an approximate evolution equation.
\item If the constraint equation (\ref{evolv}) is to be regarded as expressing evolution with respect
to an internal `time' $\phi$, how is its solution, $\Psi(\phi,v)$ to be interpreted? A statement to
the effect that $\vert\Psi(\phi,v)\vert^2\rmd v$ is the probability of finding the volume to have
a value between $v$ and $v+\rmd v$ at `time' $\phi$ (or a similar statement that replaces $\rmd v$
with a more appropriate measure if necessary) as in standard quantum mechanics will not do, because
the scalar field is a physical quantity that has no definite value until it is measured; $\phi$
is not an external parameter with the heraclitian property of `setting the conditions' for a
measurement of $v$.  For the same reason, the `Heisenberg-picture' operator (\ref{vofphi0}) cannot
be construed as representing the volume at `time' $\phi_0$.  Nor will it be possible to interpret
$\vert\Psi(\phi,v)\vert^2$ in terms of a joint probability for obtaining the pair of values
$\phi$ and $v$ from simultaneous measurements of the scalar field and the volume.  The reason is that
there are not enough physically meaningful quantities available to be measured.  Classically, the
4-dimensional kinematical phase space, with coordinates ($v$, $p_v$, $\phi$, $p_\phi$) is
reduced by the constraint to a 2-dimensional physical phase space of distinct gauge orbits, with
a single pair of conjugate coordinates.  Quantum-mechanically, this means that only one independent
quantity is available to be measured.  Equivalently, one cannot define two independent operators
$\hat{v}$ and $\hat{\phi}$ acting in the physical Hilbert space; there is only one physical
degree of freedom, corresponding, for example, to the `complete observable' $\hat{V}(0)$.\footnote{The
same point can be phrased in terms of probability measures: if $\vert\Psi(\phi,v)\vert^2$ is to
be interpreted as a joint probability density, it must be normalized with respect to a probability
measure $\rmd\mu(\phi,v)$ on a 2-dimensional configuration space.  However, the physical Hilbert
space to which a solution of (\ref{evolv}) belongs is something like $L^2(\mathbb{R},\rmd\mu(v))$,
the details depending on the quantization scheme, with a $\phi$-independent inner product.\label{measurenote}}
\end{enumerate}
It can be argued \cite{page} that even in non-relativistic quantum mechanics the external time
$t$ is irrelevant to physics.  After all, a student who investigates the motion of a pendulum
has only a set of recorded position measurements and stopwatch readings to work with in any
subsequent analysis.  For generally-covariant systems, the possibility of describing physics
entirely in terms of correlations between Dirac observables has been studied in some detail in
\cite{gambinietal}. From a purely operational point of view, it is no doubt true that substantive
physics deals only with correlations between measured quantities, but this
seems to offer an impoverished account of the world as it is actually experienced.  A student
who forgets to bring a stopwatch to the lab is not thereby prevented from seeing a pendulum
swing;  and a statement such as ``the stopwatch read 5s some three seconds after it read 2s'' seems
not to be entirely vacuous, regardless of whether it is checked for accuracy with the use
of a further clock. Moreover, while the notion of time may be scarcely
less nebulous in Newtonian mechanics than it notoriously appeared to Augustine\cite{augustine},
classical general relativity provides a concrete meaning for $t$ as the proper time that elapses
along an observer's worldline, even though it affords no experimental procedure for determining
the actual values of $t$.  That is, it seems meaningful to regard the readings of a physical clock
as supplying an estimate (more or less reliable according to the quality of manufacture) of the
intervals of geometrical proper time that elapse along its worldline, even though we have no
experimental means of checking that this is actually so.

To summarize, the interpretation of the wavefunction $\Psi(\phi,v)$ that satisfies the constraint
equation (\ref{evolv}) presents two related difficulties.  On the one hand, there is no `heraclitian'
time variable that would allow us to make sense of the theory in terms of our familiar sense of the
passage of time.  The volume $v$ and scalar field $\phi$ are physical quantities which, in principle,
can be measured. (That is, they are apparently the same sort of thing as electromagnetic fields, and
only the practical difficulty of constructing apparatus that couples to them stands in the way of
making such measurements.)  It ought to be possible to formulate questions such as ``given that I
have just determined the volume and scalar field to be $v_0$ and $\phi_0$, what is the probability that
I will find them to be $v_1$ and $\phi_1$ in an hour's time?'', but the theory as it stands does not
admit such questions.  On the other hand, if we rule that questions of this kind are inadmissible,
and confine ourselves to studying correlations between measured values of $v$ and $\phi$, we find that
this cannot be done either, because there are not enough Dirac observables to be measured.

Of course, a more comprehensive theory will afford more Dirac observables to be correlated, but that
is not of much help.  For example, a theory with an extra scalar field deals, apparently, with three
measurable quantities, $v$, $\phi$ and $\psi$, say, but yields only two independent Dirac observables,
say $V(\psi_0)$ and $\Phi(\psi_0)$, constructed along the lines of (\ref{vofphi0}).  We still face the
problem that the number $\psi_0$ cannot be interpreted as the result of a measurement of $\psi$, because
there is no corresponding Dirac observable available to be measured.  Clearly, the same will apply
to theories with more than one geometrical variable, such as the Bianchi I model studied in
\cite{martin}.

For the simple model studied here, the wavefunction $\Psi(\phi,v)$ computed in \cite{aps1,aps2} \textit{looks}
as if it describes a correlation between $v$ and $\phi$, and seems to be peaked along a classical trajectory
$(\phi(t), v(t))$. The problem is that this appearance is at variance with the number of observables to hand
(or, equivalently, with the dimension of the configuration space on which the wavefunction is defined---see
footnote \ref{measurenote}.)

In the remainder of this paper, we describe a possible solution to these difficulties, based on a view of
time evolution proposed in \cite{le}. It is useful to suppose that, as in standard quantum mechanics,
the wavefunction $\Psi(\phi,v)$ describes the state of a system from the point of view of an observer
external to the system itself.  In our case, the hypothetical observer is external to the entire model universe,
and any sense that this observer might have of a `passage of time' is quite separate from what passes inside
the universe.  The external observer has the possibility of determining the value of only a single
Dirac observable; this one value completely specifies the state of the universe--which means its entire
history.  The difficulties identified above do not concern the information accessible to the external
observer, which is delivered by the wavefunction $\Psi(\phi,v)$ according to the usual quantum-mechanical
rules, but rather what might be observed by an observer internal to the universe.  To assess the latter, it
is necessary to include in our model a description of relevant features of the physical system that does
the internal observing:  at a minimum, the clock from which this system gains its sense of time. By doing
this, as we shall illustrate, it is possible to alleviate both of the difficulties.  On the one hand, we
introduce a genuine `heraclitian' time $\tau$, corresponding to the arc-length of the observer's worldline,
\textit{not} to readings of a physical clock. Time evolution with respect to $\tau$ is described by a
standard Schr\"{o}dinger equation, precisely analogous to (\ref{schrod}), though in practice we will deal
with the corresponding Heisenberg-picture operators $\hat{V}(\tau)$ and $\hat\Phi(\tau)$. (Unruh and Wald
\cite{uw} argue that the problems of interpretation are not alleviated by the introduction of observers,
but the role they envisage for these observers is different from the one used here.) On the other hand,
the lack of a Dirac observable corresponding to the reading of the physical clock can be understood by
considering that this clock is internal to the observing apparatus, and thus \textit{in principle}
unobservable by the observer from whose point of view the evolution is described.

Clearly, the idea of incorporating an observer's clock into our model is in some way akin to the notion
of a material reference frame, which has been widely studied (see, for example
\cite{ish,kuchar,rovref,brown,montani,montani09,montani04,giesel,amemiya,thiemannphan}).  We think that the implementation of
this general idea described below differs in important respects from others to be found in the
literature, and will return to this point in section \ref{discussion}.
\section{Conventional quantization\label{conqu}}
As a rule, cosmologists do not find it necessary to include the energy content of their observing
apparatus explicitly in the Friedmann equation.  Correspondingly, we seek to modify the cosmological
model defined by (\ref{action1}) by introducing a `test clock', which will serve to reveal the time
evolution of the volume and scalar field, while disturbing the Hamiltonian constraint to a negligible
extent.  In this section, we consider the quantum theory of such a model, using a conventional quantization
scheme similar, though not identical, to the `Wheeler-de-Witt' scheme described in \cite{aps1,aps2,acs,acsrev},
within which most of the analysis can be achieved exactly and explicitly.  In section \ref{loopqu}, we
will consider a `polymer' quantization scheme of the type employed in loop quantum cosmology, where we
will need to resort to approximations.  In either case, we may expect (na\"{i}vely, on the basis of an
`energy-time uncertainty relation') that restricting the energy of the clock should place some limit on
the resolution with which the value of a time-dependent Dirac observable can be determined, and we will
investigate this issue in some detail.
\subsection{Homogeneous cosmology with a test clock\label{hctc}}
Classically, consider a small clock, whose internal workings are described by a Lagrangian $\ell(\lambda)$,
localized on the worldline\footnote{A more rigorous derivation than we attempt here would, amongst other refinements,
consider the clock to be localized in a region somewhat larger than its Schwarzschild radius, but
we assume that such refinements are inessential to the issue of time evolution that is our main concern.}
$x^\mu(\lambda)$, parametrized by its arc length $\lambda$. Its contribution to the action has, in general,
the form
\begin{eqnarray}
S_\mathrm{clock}&=&\int\rmd^4x\sqrt{\vert g(x)\vert}\int\rmd\lambda \ell(\lambda)\delta(x,x(\lambda))
\label{clockaction}\\
&=&\int\rmd\lambda \ell(\lambda)
\end{eqnarray}
where $\delta$ is a covariant delta function, with the property $\int\rmd^4x\sqrt{\vert g(x)\vert}\delta(x,x')f(x)=f(x')$.
In principle, the coordinates $x^\mu(\lambda)$ are dynamical variables, but in order to construct a simple
cosmological model, we remove these degrees of freedom, along with most of those in $g_{\mu\nu}$, by taking
the world line to be that of a comoving observer in a FRW universe. Then the arc length $\lambda$ coincides
with $t$ in (\ref{propertime}) and the total action becomes
\begin{equation}
S=\int\rmd sN(s)^{-1}\left[-\frac{1}{24\pi G}\frac{(\partial_sv)^2}{v}+\frac{1}{2}v(\partial_s\phi)^2
\vphantom{\frac{(\partial_sv)^2}{v} }+N(s)^2\ell(t(s))\right].
\label{action}
\end{equation}
The Hamiltonian constraint now reads
\begin{equation}
H_0+h=-6\pi G vp_v^2+\frac{1}{2}v^{-1}p_\phi^2+h=0
\end{equation}
where $h$ is the Legendre transform of $\ell$, and is intended to be very small compared with the matter
term in $H_0$, in the same sense that a silicon chip on board WMAP is small compared with the total
energy content of the visible universe. Since the action (\ref{action}) is supposed to apply to a
homogeneous universe, it might be more consistent to regard the last term as arising from a space-filling
congruence of clocks.  However, if $h$ is small enough, then it seems very likely that, say, a spherically
symmetric cosmology, with the clock definitely localized on a single world line, but departing only to this
tiny extent from a genuinely homogeneous one, would be described by the action (\ref{action}) with negligible
error.  What is essential for our purpose is that, because the clock (or each clock) is localized on a
single worldline, $h$ is independent of the cosmological variables $(v, p_v, \phi, p_\phi)$. Having obtained
the constraint, we will dispense with the arbitrary coordinate $s$, and deal only with the proper time $t$ or,
equivalently, set $N(s)=1$ so that $s$ and $t$ coincide. \textit{Up to the choice of an origin}, $t$ has
a clear physical interpretation as the geometrical proper time that elapses along the comoving observer's worldline.

We will not be specific about the microscopic constitution of the clock.  Suppose that $h$ depends on several
microscopic phase-space variables, which we denote collectively by $\bm{x}$, and that $\bar{\bm{x}}(\bm{x},t)$
is the phase-space trajectory (the solution of $\partial_t\bar{\bm{x}}=\{\bar{\bm{x}},h\}$) that passes
through $\bm{x}$ at, say, $t=0$. The reading of the clock is some function $r(\bm{x})$, and we define the
function $t_0(\bm{x})$ as the solution of
\begin{equation}
r(\bar{\bm{x}}(\bm{x},t_0))=0.\label{t0def}
\end{equation}
(The clock is fit for purpose only if this equation has a unique solution.)

 For a given state of the clock, the
quantity $\tau=t-t_0$ has a classical interpretation as the interval of proper time that has elapsed since the
clock read 0.  This fiducial event provides a physical meaning for the origin $\tau=0$.  By contrast, the
arc-length parameter $t$ is defined by (\ref{propertime}) only up to an arbitrary constant, because a change
$\delta{N}(s)$ in the undetermined lapse function changes $t$ by an additive constant.  Fixing an origin
for $\tau$ is the only essential role of the physical clock.  It is not hard to show that
\begin{equation}
\{t_0,h\}=-1
\end{equation}
and this is the only property of the clock that will matter.

Finally, let $\bm{w}$ denote collectively the phase-space variables that appear in $H_0$ (they are
$(v,p_v,\phi,p_\phi)$ in the present example, but the following result is general).  For some quantity
$f(\bm{w})$, let $\bar{f}(\bm{w},t)$ be the solution of
\begin{equation}
\partial_t\bar{f}=\{\bar{f},H_0\}
\end{equation}
with the initial condition $\bar{f}(\bm{w},0)=f(\bm{w})$. Then
\begin{equation}
F(\tau):=\bar{f}(\bm{w},t_0+\tau)\label{Foftau}
\end{equation}
is classically the value of $f$ at a proper time $\tau$ after the clock read 0. Since $h$ and $t_0$ are
independent of $\bm{w}$, and the vector field $\{\bm{\cdot}\,,h\}$ is a linear differential operator, we easily find
\begin{equation}
\{F(\tau),h\}=\partial_\tau F(\tau)\{t_0,h\}=-\{F(\tau),H_0\},\label{classDiracF}
\end{equation}
so for each $f$ and $\tau$, $F(\tau)$ commutes with the constraint $H=H_0+h$.  Thus, $F(\tau)$ gives a
1-parameter family of Dirac observables\footnote{The construction of these observables is somewhat similar to
the construction of `evolving constants of the motion' proposed by Rovelli \cite{rovtime}, as mentioned above,
and elaborated by others, \textit{but it is not the same construction}. We think the difference is
important, and discuss it in detail in section \ref{discussion}.}, and obeys the equation of motion
\begin{equation}
\partial_\tau F(\tau)=\{F(\tau),H_0\}.\label{Feom}
\end{equation}

For the model at hand, it proves convenient to define $y=vp_v$.  Then the Hamilton equations are
\begin{eqnarray}
\partial_t \bar{v}&=&-12\pi G\bar{y}\\
\partial_t\bar{y}&=&H_0\\
\partial_t\bar{\phi}&=&\bar{v}^{-1}p_\phi\\
\partial_t p_\phi&=&0
\end{eqnarray}
and they have the solutions
\begin{eqnarray}
\bar{v}(t)&=&v-12\pi G y t-6\pi G H_0t^2\\
\bar{y}(t)&=&y+H_0t\\
\bar{\phi}(t)&=&\phi+\frac{1}{\sqrt{12\pi G}}\ln\left(\frac{v-a_-t}{v-a_+t}\right)\label{phibar}
\end{eqnarray}
where $a_{\pm}=6\pi Gy\pm\sqrt{3\pi G}p_\phi$ arise from the factorization $\bar{v}(t)=v^{-1}(v-a_+t)(v-a_-t)$.
We can now define a collection of basic Dirac observables
\begin{eqnarray}
V&=&v-12\pi G y t_0-6\pi G H_0t_0^2\\
Y&=&y+H_0t_0\\
\Phi&=&\phi+\frac{1}{\sqrt{12\pi G}}\ln\left(\frac{v-a_-t_0}{v-a_+t_0}\right)\\
P_\Phi&=&p_\phi.
\end{eqnarray}
Of course, the form of the Hamiltonian is preserved
\begin{eqnarray}
H_0&=&-6\pi G v^{-1}y^2 +\frac{1}{2}v^{-1}p_\phi^2\nonumber\\
&=&-6\pi G V^{-1}Y^2 +\frac{1}{2}V^{-1}P_\Phi^2,
\end{eqnarray}
the gauge-invariant variables $V$ and $Y$ inherit the Poisson-bracket relations satisfied by $v$ and $y$,
namely
\begin{equation}
\{V,Y\}=V,\qquad\{V,H_0\}=-12\pi G Y,\qquad\{Y,H_0\}=H_0,
\end{equation}
and the equations of motion (\ref{Feom}) have the obvious solutions
\begin{eqnarray}
V(\tau)&=&V-12\pi G Y \tau-6\pi G H_0\tau^2\label{cvoftau}\\
Y(\tau)&=&Y+H_0\tau\\
\Phi(\tau)&=&\Phi+\frac{1}{\sqrt{12\pi G}}\ln\left(\frac{V-A_-\tau}{V-A_+\tau}\right)\label{cphioftau}\\
P_\Phi(\tau)&=&p_\phi,
\end{eqnarray}
with $A_{\pm}=6\pi GY\pm\sqrt{3\pi G}P_\Phi$.\\

\subsection{Quantum Dirac observables}\label{DO}
Formally, it is easy to construct quantum-mechanical versions of the classical Dirac observables given
above. Introduce operators $\hat{H}_0$, $\hat{f}$, $\hat{h}$ and $\hat{t}_0$, which have, in particular,
the commutators
\begin{equation}
[\hat{h},\hat{t_0}]=\rmi\hbar,\qquad[\hat{H}_0,\hat{h}]=[\hat{f},\hat{h}]=[\hat{H}_0,\hat{t}_0]=[\hat{f},\hat{t}_0]=0.
\end{equation}
Then it is straightforward to verify that the operator
\begin{equation}
\hat{F}(\tau):=\exp\left[\frac{\rmi}{\hbar}\hat{H}_0(\hat{t_0}+\tau)\right]\hat{f}
\exp\left[-\frac{\rmi}{\hbar}\hat{H}_0(\hat{t_0}+\tau)\right]\label{diracF}
\end{equation}
commutes with the constraint,
$
[\hat{F}(\tau),\hat{H}_0+\hat{h}]=0,
$
and obeys the Heisenberg equation of motion
\begin{equation}
\partial_\tau\hat{F}(\tau)=\frac{\rmi}{\hbar}[\hat{H}_0,\hat{F}(\tau)].\label{qeom}
\end{equation}
In line with previous notation, we write $\hat{F}:=\hat{F}(0)$.

There is no guarantee \textit{a priori} that the solutions to (\ref{qeom}) will be simply related to the
classical expressions (\ref{cvoftau})-(\ref{cphioftau}).  We will insist on the commutation relations
\begin{equation}
[\hat{v},\hat{p}_v]=[\hat{\phi},\hat{p}_\phi]=\rmi\hbar,
\end{equation}
with all other commutators amongst $\hat{v},\hat{p}_v,\hat\phi$ and $\hat{p}_\phi$ vanishing, and choose
for the Hamiltonian the operator ordering
\begin{equation}
\hat{H}_0=-6\pi G\hat{p}_v\hat{v}\hat{p}_v+\frac{1}{2}\hat{v}^{-1}\hat{p}_\phi^2.
\end{equation}
Then define
\begin{equation}
\hat{y} =\frac{1}{2}(\hat{v}\hat{p}_v+\hat{p}_v\hat{v}),
\end{equation}
which also implies $\hat{p}_v=\frac{1}{2}(\hat{v}^{-1}\hat{y}+\hat{y}\hat{v}^{-1})$ and
leads to the closed set of commutation relations
\begin{equation}
[\hat{v},\hat{y}]=\rmi\hbar\hat{v},\quad[\hat{H}_0,\hat{v}]=\rmi\hbar12\pi G\hat{y},\quad
[\hat{H}_0,\hat{y}]=-\rmi\hbar\hat{H}_0.
\end{equation}
We then indeed find
\begin{eqnarray}
\hat{v}(t)&=&\hat{v}-12\pi G\hat{y}t-6\pi G\hat{H}_0t^2\\
\hat{y}(t)&=&\hat{y}+\hat{H}_0t,
\end{eqnarray}
from which $\hat{V}$ and $\hat{Y}$ can be defined by replacing $t$ with $\hat{t}_0$, and
\begin{eqnarray}
\hat{V}(\tau)&=&\hat{V}-12\pi G \hat{Y} \tau-6\pi G \hat{H}_0\tau^2\label{vhatoftau}\\
\hat{Y}(\tau)&=&\hat{Y}+\hat{H}_0\tau.
\end{eqnarray}
Since $\hat{p}_\phi$ commutes with $\hat{H}_0$, we can also identify $\hat{P}_\Phi(\tau)=\hat{P}_\Phi=\hat{p}_\phi$.
The Dirac observable associated with $\phi$ is more difficult to write down, as might be expected
from the nonlinearity of the classical expression (\ref{phibar}). However, this expression shows that
$\rme^{\tilde{\alpha}\bar{\phi}(t)}\bar{v}(t)$, with $\tilde{\alpha}=\sqrt{12\pi G}$, is quadratic in $t$,
and it can be verified, by taking repeated commutators with $\hat{H}_0$, that
\begin{equation}
\rme^{\tilde{\alpha}\hat{\Phi}}\hat{V}:=\rme^{\rmi\hat{H}_0\hat{t}_0/\hbar}\rme^{\tilde{\alpha}\hat{\phi}}\hat{v}
\rme^{-\rmi\hat{H}_0\hat{t}_0/\hbar}
=\rme^{\tilde{\alpha}\hat{\phi}}\left[
\hat{v}+\hat{a}\hat{t}_0+\frac{1}{2}\hat{b}\hat{t}_0^2\right]\label{ephiv}
\end{equation}
where
\begin{eqnarray*}
\hat{a}&=&\tilde{\alpha}\hat{p}_\phi-\tilde{\alpha}^2\left(\hat{y}+\frac{\rmi\hbar}{2}\right)\\
\hat{b}&=&\left(\hat{a}+\tilde{\alpha}^2\hat{y}\right)\hat{v}^{-1}\hat{a}-\tilde{\alpha}^2
\hat{H}_0.
\end{eqnarray*}
Sadly, we have not found any useful closed-form expression for $\hat{\Phi}(\tau)$.
\subsection{Physical Hilbert space}
Take a kinematical vector space of wavefunctions $\Psi(v,p_\phi,h)$, in which operators $\hat{v}$,
$\hat{p}_\phi$ and $\hat{h}$ act by multiplication, while
\begin{equation}
\hat{p}_v\Psi=-\rmi\hbar\frac{\partial\Psi}{\partial v},\qquad
\hat\phi\Psi=\rmi\hbar\frac{\partial\Psi}{\partial p_\phi},\qquad
\hat{t}_0\Psi=-\rmi\hbar\frac{\partial\Psi}{\partial h}.
\end{equation}
The Hamiltonian $\hat{H}_0$ is
\begin{equation}
\hat{H}_0\Psi=\left[\frac{1}{2}\alpha^2\partial_vv\partial_v+\frac{1}{2}v^{-1}p_\phi^2\right]\Psi,
\end{equation}
where $\alpha^2=\tilde{\alpha}^2\hbar^2=12\pi G\hbar^2$, and the constraint reads
\begin{equation}
\left[\frac{\alpha^2}{2}v\partial_vv\partial_v+\frac{1}{2}p_\phi^2+vh\right]\Psi(v,p_\phi,h)=0.
\end{equation}
The change of variables
\begin{equation}
\nu=\frac{2\rmi p_\phi}{\alpha},\qquad z=\frac{2\sqrt{2h}}{\alpha}v^{1/2}\label{nuz}
\end{equation}
converts this into a standard form of Bessel's equation
\begin{equation}
\left(z\partial_zz\partial_z-\nu^2+z^2\right)\Psi=0,\label{bessel}
\end{equation}
and we find
\begin{eqnarray}
\hat{v}\Psi&=&\frac{\alpha^2}{8h}z^2\Psi\label{vhat}\\
\hat{y}\Psi&=&-\frac{\rmi\hbar}{2}\left(z\partial_z+1\right)\Psi\\
\hat{t}_0\Psi&=&-\rmi\hbar h^{-1}\left(\frac{1}{2}z\partial_z+h\partial_h\right)\Psi\label{t0hat}\\
\hat{H}_0\Psi&=&\frac{h}{z^2}\left(z\partial_zz\partial_z-\nu^2\right)\Psi.
\end{eqnarray}

The constraint equation (\ref{bessel}) is solved by
\begin{equation}
\Psi(v,p_\phi,h)=\psi(p_\phi,h)\mathcal{C}_\nu(z),
\end{equation}
where $\mathcal{C}_\nu(z)$ is some Bessel function, and $\psi(p_\phi,h)$ is arbitrary. (More generally,
the solution is a linear combination of two independent Bessel functions of order $\nu$, but these two sectors are
not mixed by any operator of interest, and we assume it is sufficient to keep just one.)
Clearly, the physical Hilbert space is the space of functions $\psi(p_\phi,h)$, equipped with a suitable
inner product. The operators (\ref{vhat})-(\ref{t0hat}) corresponding to `partial observables', do not,
of course, have any well-defined action in this space, but the Dirac observables do, and we easily verify that
\begin{eqnarray}
\hat{V}\psi&=&-\frac{1}{2}\left(\alpha^2\partial_hh\partial_h+h^{-1}p_\phi\right)\psi\label{vhath}\\
\hat{Y}\psi&=&\rmi\hbar h^{1/2}\partial_hh^{1/2}\psi\\
\hat{H}_0\psi&=&-h\psi.\label{H0h}
\end{eqnarray}
The operator $\hat{\Phi}$ is again harder to deal with.  Note in particular that, although its conjugate
momentum $\hat{P}_\Phi$ is just multiplication by $p_\phi$, we cannot identify
$\hat{\Phi}=\rmi\hbar\partial/\partial p_\phi$, since it does not commute with $\hat{H}_0$. However, a
well-defined action of the Dirac observable (\ref{ephiv}) can be found as follows.  Acting on the Bessel
function, we have
\begin{equation}
\rme^{\tilde{\alpha}\hat\phi}\mathcal{C}_\nu(z)=\rme^{-2\partial/\partial\nu}\mathcal{C}_\nu(z)
=\mathcal{C}_{\nu-2}(z).
\end{equation}
Then, repeated use of the standard recurrence relations that connect $\mathcal{C}_{\nu\pm n}(z)$ and
their derivatives suffices to verify that
\begin{equation}
\rme^{\tilde{\alpha}\hat\Phi}\hat{V}\psi(\nu,h)=\frac{\alpha^2}{8h}\left[\nu(\nu-2)-4(\nu-1)h\partial_h
+4h\partial_hh\partial_h\right]\psi(\nu-2,h),\quad\label{phiop}
\end{equation}
where, as a shorthand, we use $\psi(\nu,h)$ to mean the same as $\psi(p_\phi,h)$ via (\ref{nuz}).

Finally, make the change of variable
\begin{equation}
h=\epsilon\rme^{\alpha\eta}\label{h2eta}
\end{equation}
where $\epsilon$ is a positive constant that we will later choose to be close to $\langle\hat{h}\rangle$.
We find
\begin{eqnarray}
\hat{V}&=&-\frac{1}{2\epsilon}\rme^{-\alpha\eta}\left(\partial_\eta^2+p_\phi^2\right)\label{vhateta}\\
\hat{Y}&=&\frac{\rmi\hbar}{\alpha}\rme^{-\alpha\eta/2}\partial_\eta\rme^{\alpha\eta/2}\label{yhateta}\\
\hat{H}_0&=&-\epsilon\rme^{\alpha\eta}=:-\hat{h}.\quad\label{heta}
\end{eqnarray}
Choosing the inner product
\begin{equation}
(\psi_1,\psi_2)=\int_{-\infty}^\infty\rmd p_\phi\int_{-\infty}^\infty\rmd\eta\,\rme^{\alpha\eta}\,
\bar{\psi}_1(p_\phi,\eta)\psi_2(p_\phi,\eta),
\end{equation}
so $\mathcal{H}_\mathrm{phys}=L^2(\mathbb{R}^2,\rmd p_\phi\rme^{\alpha\eta}\rmd\eta)$, these three operators
and a fourth one derived from (\ref{phiop}) are symmetric\footnote{We do not address the issue of
self-adjointness; it is sufficient for our illustrative purposes that these operators have well defined
matrix elements between Gaussian wavefunctions.}. It is worth noting that this quantization scheme differs
from the Wheeler-de-Witt scheme described in \cite{aps2} not only by the addition of the clock variable,
but also because the inner product is chosen to make the generator of displacements in $\tau$ self-adjoint,
rather than the generator of displacements in $\phi$.
\subsection{Viability of the test clock\label{viab1}}
A wavefunction $\psi(p_\phi,\eta)$ in the physical Hilbert space specifies
the state of a system consisting of two physical objects, the volume and the scalar field, represented by
the pairs of Dirac observables $(\hat{V},\hat{Y})$ and $(\hat\Phi,\hat{P}_\Phi)$ respectively, but the clock
does not feature as an independent object;  it has been eliminated by solving the constraint.  As discussed
in the previous section, we interpret this state of affairs by supposing that the clock is internal to an
observer (or, less anthropomorphically, an observing apparatus) who is thereby debarred in principle from
observing it.  The remaining Dirac observables describe the volume and scalar field from the point of view
of this observer, and evolve precisely as in standard quantum mechanics with respect to the `heraclitian'
time $\tau$.

If this is to make sense, it must be possible to find states in which the clock serves to reveal the time
evolution of the universe, while perturbing it to a negligible extent.  One way of ensuring this would be
to model the clock as an system whose energy is bounded.  Thus, we might attempt to realize the Dirac
observables (\ref{vhath})-(\ref{H0h}) as self-adjoint operators on, say, $\mathcal{H}_\mathrm{phys}=
L^2(\mathbb{R}\times[0,\epsilon],\rmd p_\phi\,\rmd h)$.  It is technically easier to retain our earlier choice
$\mathcal{H}_\mathrm{phys}=L^2(\mathbb{R}^2,\rmd p_\phi\rme^{\alpha\eta}\rmd\eta)$, on which the clock's
energy $\hat{h}$ is positive (see (\ref{h2eta})), and restrict attention to \textit{states} in which this
energy is small.  It is not obvious that one of these choices is physically less reasonable than the other.
Given the actual state of our world, the probability of finding the energy content of, say, a small alarm
clock to be a significant fraction of the energy content of the visible universe is (presumably) extremely
small.  Depending on details of the clock's construction, a state in which that probability is appreciable
might not be impossible in principle, though it would bear very little resemblance to the state that we
actually have.

In the model universe, we thus want to find states in which the probability of finding the clock's energy to be a
significant fraction of the energy carried by the scalar field is very small; that is, we need both
the expectation value $\epsilon_\rmc:=\langle\hat{h}\rangle$ and the variance
$(\delta h)^2:=\langle(\hat{h}-\epsilon_\rmc)^2\rangle$ to be very small. Then the following issue arises.
A simple-minded argument would suggest that, if $\delta h$ is vanishingly small, one has in effect an eigenstate
of $\hat{H}_0$, say $\hat{H}_0\vert\epsilon_\rmc\rangle=-\epsilon_\rmc\vert\epsilon_\rmc\rangle$.  In that state,
the expectation value of any commutator $\langle\epsilon_\rmc\vert[\hat{F},\hat{H}_0]\vert\epsilon_\rmc\rangle$
vanishes, and no time dependence
should be apparent.  This argument is in fact too simple-minded, because $\vert\epsilon_\rmc\rangle$ is not a
normalizable state in $\mathcal{H}_\mathrm{phys}$; what \textit{does} happen is that the Dirac observables
$\hat{V}(\tau)$ and $\hat{Y}(\tau)$ have very large variances when $\delta h$ is very small. Their probability
distributions cannot be sharply peaked along some quasi-classical trajectory, and in that sense they are not
well-defined time-dependent observable quantities.  The important issue, then, is whether a useful compromise
exists, such that the clock contributes, with high probability, only a negligible amount to the total energy
of the universe, while the time-dependent observables remain reasonably well defined.

In any state, the expectation value of the time-dependent volume operator (\ref{vhatoftau}) is
\begin{equation}
V(\tau):=\langle\hat{V}(\tau)\rangle=V+3\mathcal{H}V\tau+\frac{\alpha^2}{\hbar^2}\epsilon_\rmc\tau^2,
\label{exvoftau}
\end{equation}
where
\begin{equation}
V:=\langle\hat{V}\rangle\qquad\hbox{and}\qquad\mathcal{H}:=-\frac{\alpha^2}{3\hbar^2V}\langle\hat{Y}\rangle
\label{vandH}
\end{equation}
are the volume and Hubble parameter when the clock reads 0. (We emphasize that $\hat{V}(\tau)$ is the volume when
an interval $\tau$ of geometrical proper time has elapsed since that fiducial event:  it \textit{cannnot} be
interpreted as the volume when the clock reads $\tau$.) The first two terms in $V(\tau)$ follow the classical
solution in the absence of the clock.  When $\tau$ is large enough, we have
$V(\tau)\approx (\alpha^2\epsilon_\mathrm{c}/\hbar^2)\tau^2$, corresponding to a scale factor
$a(\tau)\propto\tau^{2/3}$.  In this situation, the energy density of the scalar field has become so dilute
that the energy content of the fiducial cell is dominated by the clock, which in this model is equivalent to
pressureless matter (see the remark following (\ref{action})).  At that point, the clock has ceased to qualify
as a `test' clock, but in a suitably chosen state, this will occur too far in the future to be a practical
concern.

Consider a state of the form
\begin{equation}
\psi(p_\phi,\eta)=\chi(p_\phi)\rme^{-\Delta\eta^2+\rmi q\eta},\label{psiofeta}
\end{equation}
where $\chi(p_\phi)$ is sufficiently sharply peaked at some value that $p_\phi$ in (\ref{vhateta}) can be
taken as just that value, which we call simply $p_\phi$.  The wavenumber $q$ is real; an imaginary part corresponds
to the constant $\epsilon$ in (\ref{h2eta}).  The clock's energy is
\begin{equation}
\epsilon_\rmc=\langle\hat{h}\rangle=\epsilon\rme^{3\alpha^2/4\Delta}
\end{equation}
and we define
\begin{equation}
E_\phi:=\frac{1}{2}V^{-1}p_\phi^2,
\end{equation}
which is the total scalar-field energy contained in the volume $V$. Then the expectation value of $\hat{V}$
turns out to be
\begin{equation}
V=\frac{1}{2}\epsilon_\rmc^{-1}\rme^{\alpha^2/2\Delta}\left[\frac{1}{2}\Delta+q^2-p_\phi^2\right],
\end{equation}
which can be rearranged to yield
\begin{equation}
q^2=2VE_\phi\left[1+\frac{\epsilon_\rmc}{E_\phi}\,\rme^{-\alpha^2/2\Delta}\right]-\frac{1}{2}\Delta,
\end{equation}
while the Hubble parameter defined by (\ref{vandH}) is
\begin{equation}
\mathcal{H}=-\frac{\alpha^2}{3\hbar^2V}\langle{\hat{Y}}\rangle=\frac{\alpha q}{3\hbar V}.\label{hubble}
\end{equation}

For the variances of the three operators (\ref{vhateta})-(\ref{heta}), we find
\begin{eqnarray}
\left(\frac{\delta V}{V}\right)^2&=&\left(\rme^{\alpha^2/2\Delta}-1\right)\nonumber\\&&+\rme^{3\alpha^2/2\Delta}\left[
\frac{\Delta}{\epsilon_\rmc V}\frac{E_\phi}{\epsilon_\rmc}\left(1+\frac{\epsilon_c}{E_\phi}\rme^{-\alpha^2/2\Delta}
\right)
\right.+\frac{1}{2}\frac{\alpha^2}{\epsilon_\rmc V}\frac{E_\phi}{\epsilon_\rmc}
\left(1+\frac{\epsilon_c}{2E_\phi}\rme^{-\alpha^2/2\Delta}\right)\left.+\frac{1}{64}\left(\frac{\alpha^2}{\epsilon_\rmc V}\right)^2-\frac{1}{8}\left(\frac{\Delta}{\epsilon_\rmc V}
\right)^2\right]\nonumber\\&&\\
\left(\frac{\delta Y}{Y}\right)^2&=&\frac{\Delta}{2q^2}\\
\left(\frac{\delta h}{h}\right)^2&=&\rme^{\alpha^2/2\Delta}-1.
\end{eqnarray}
If the clock energy $\epsilon_\rmc$ is to contribute a negligible amount to the constraint, we need
\begin{equation}
\frac{\epsilon_\rmc}{E_{\phi}}\ll 1,\label{epsoverE}
\end{equation}
and if the first term in $(\delta V/V)^2$ is to be small, we also need
\begin{equation}
\frac{\alpha^2}{2\Delta}\ll1.\label{alphaoverdelta}
\end{equation}
With these approximations, we get
\begin{equation}
\left(\frac{\delta V}{V}\right)^2\approx\frac{\alpha^2}{2\Delta}
+\frac{\Delta}{\epsilon_\rmc V}\frac{E_\phi}{\epsilon_\rmc}-\frac{1}{8}\left(\frac{\Delta}{\epsilon_\rmc V}\right)^2,
\label{deltavapprox}
\end{equation}
and if \textit{this} is to be small, we finally need
\begin{equation}
\frac{\Delta}{\epsilon_\rmc V}\ll\frac{\epsilon_\rmc}{E_\phi}.\label{deltaoverepsilonv}
\end{equation}
If these conditions are met, then $q^2\approx2VE_\phi$ and $(\delta Y/Y)^2\approx (\Delta/4\epsilon_\rmc V)
(\epsilon_\rmc/E_\phi)$ is small. Also, the Hubble parameter given by (\ref{hubble}) becomes
\begin{equation}
\mathcal{H}^2\approx\left(\frac{\alpha}{3\hbar V}\right)^22VE_\phi=\frac{8\pi G}{3}\frac{E_\phi}{V}
=:\frac{8\pi G}{3}\rho_\phi
\end{equation}
which is the classical Friedmann equation.

With the approximations (\ref{epsoverE}), (\ref{alphaoverdelta}) and (\ref{deltaoverepsilonv}), the
variance of the time-dependent volume operator gives
\begin{equation}
\left(\frac{\delta V(\tau)}{V}\right)^2\approx\frac{\alpha^2}{2\Delta}\left[1-\frac{\epsilon_c}{4E_\phi}
(3\mathcal{H}\tau)^2\right]^2
+\frac{\Delta}{\epsilon_\rmc V}\frac{E_\phi}{\epsilon_\rmc}
\left[1+\frac{\epsilon_c}{2E_\phi}(3\mathcal{H}\tau)\right]^2,
\end{equation}
where, on the LHS, $V$ in the denominator is $\langle\hat{V}\rangle$, not $V(\tau)$.  As expected, this
leads to the same physics as the version of the Wheeler-de-Witt equation studied in \cite{aps2}.  Namely, the singularity
occurs when $3\mathcal{H}\tau\approx -1$, at which point $\delta V(\tau)$ is substantially unchanged from
its value at $\tau=0$.

The compromise needed for the `test clock' to make sense consists in satisfying the two inequalities
\begin{equation}
\frac{\alpha^2}{2}\ll\Delta\ll\frac{\epsilon_\rmc^2V}{E_\phi}\label{tccomp}
\end{equation}
simultaneously.  For an essentially classical state at $\tau=0$, this is indeed possible---and by a huge margin,
as might be expected.  In SI units, we have, first of all
\begin{equation}
\alpha^2=\frac{12\pi G_\mathrm{N}\hbar^2}{c^2}\approx3.1\times10^{-94}\mathrm{J\,m^3}.
\end{equation}
Although our universe is not dominated by the energy of a massless scalar field, for illustrative
purposes we take $\rho_\phi=E_\phi/V\sim 10^{-9}\mathrm{J\,m^{-3}}$ (roughly the observed current energy density) and a clock with energy
$\epsilon_\rmc=1\,\mathrm{kg}.c^2\sim 10^{17}\mathrm{J}$. Then
\begin{equation}
\frac{\epsilon_\rmc^2V}{E_\phi}=\frac{\epsilon_\rmc^2}{\rho_\phi}\sim10^{43}\mathrm{J\,m^3}.
\end{equation}
Evidently, there should be no problem finding a $\Delta$ in the range (\ref{tccomp}).

The time $\tau_\rmc$ at which the scalar field energy is so diluted that the clock begins to perturb the evolution
significantly is given by (\ref{exvoftau}) as
\begin{equation}
\tau_\rmc=\frac{\hbar^2}{\alpha^2\epsilon_\rmc}3\mathcal{H}V=\frac{2\rho_\phi V}{\epsilon_\rmc}\tau_0,
\end{equation}
where $\tau_0=(3\mathcal{H})^{-1}$ is the age of the universe at the fiducial time $\tau=0$.  If $V$ is
the size of the observable universe, say $V=10^{80}\mathrm{m^3}$, we get $\tau_\rmc\sim10^{54}\tau_0$, so
the last term in (\ref{exvoftau}) can safely be ignored for practical purposes, but that would clearly be true
for much smaller volumes too.

If $\Delta/\epsilon_\rmc V\ll\epsilon_\rmc/E_\phi$, then the last term in (\ref{deltavapprox}) is negligible,
and the sum of the remaining terms is minimized by
\begin{equation}
\Delta=\sqrt{\frac{\alpha^2\epsilon_\rmc^2V}{2E_\phi}},\label{deltamin}
\end{equation}
namely the geometric mean of the two extreme values in (\ref{tccomp}).  It is possibly interesting to
consider this compromise in the context of the Robertson-Schr\"odinger uncertainty relation \cite{schrodinger}
\begin{equation}
(\delta V)^2(\delta Y)^2\ge\frac{1}{4}\left\vert\langle[\hat{V},\hat{Y}]\rangle\right\vert^2
+\frac{1}{4}\left\vert\langle\{\hat{V}-V,\hat{Y}-Y\}\rangle\right\vert^2.
\end{equation}
With the three approximations (\ref{epsoverE}), (\ref{alphaoverdelta}) and (\ref{deltaoverepsilonv}), this
inequality is saturated, both sides being given approximately by
\begin{equation}
\frac{\hbar^2}{4}V^2\left[1+\frac{2\Delta}{\alpha^2}\,\frac{\Delta}{\epsilon_\rmc V}\,\frac{E_{\phi}}{\epsilon_\rmc}
\right],
\end{equation}
and the value (\ref{deltamin}) of $\Delta$ makes the commutator and anticommutator terms equal.
\section{Loop quantum cosmology with a test clock\label{loopqu}}
\subsection{Loop classical cosmology}
The cosmological model treated in \cite{aps1,aps2,acs,acsrev} arises, in effect, from quantization of
the classical Hamiltonian
\begin{eqnarray}
H_0&=&H_\mathrm{grav}(\nu,\sigma)+H_\mathrm{matter}(\nu,p_\phi)\\
H_\mathrm{grav}&=&\frac{\alpha^2B}{4}\nu\sin^2(2\sigma)\left[\sin\sigma\{\cos\sigma,\vert\nu\vert\}
-\cos\sigma\{\sin\sigma,\vert\nu\vert\}\right]\label{Hgrav}\\
H_\mathrm{matter}&=&B\vert\nu\vert^{-1}p_\phi^2.
\end{eqnarray}
In terms of the volume $v$ and its conjugate momentum $p_v$ defined above, the new variables are given by
\begin{equation}
\nu=2B\sgn(\nu)v,\qquad\sigma=-\frac{\sgn(\nu)}{2\hbar B}p_v
\end{equation}
where $B=\beta\hbar^{3/2}\alpha^{-3}$ and the numerical constant $\beta=2^{1/2}3^{5/4}\gamma^{-3/2}$ is
approximately equal to $48.24$, when the Barbero-Immirzi parameter $\gamma$ is given the value $0.2375\ldots$ obtained
from calculations of black-hole entropy in loop quantum gravity \cite{dom,mei}. Their Poisson bracket is
\begin{equation}
\{\sigma,\nu\}=\hbar^{-1}.\label{sigmanupb}
\end{equation}
The motivation for this Hamiltonian is explained in detail in \cite{aps1,aps2,acs,acsrev} (see also \cite{abl});
here, we note only
that the sign of the volume variable $\nu$ corresponds to the orientation of a physical co-triad relative to a
fiducial one, and that $H_0$ reduces to the original form (\ref{H0}) in the limit $\hbar\to0$.

By using the basic Poisson bracket (\ref{sigmanupb}) to evaluate those in (\ref{Hgrav}), we arrive at the Hamiltonian
\begin{equation}
H_\mathrm{LCC}=-\frac{\alpha^2B}{4}\vert\nu\vert\sin^2(2\sigma)+B\vert\nu\vert^{-1}p_\phi^2,
\end{equation}
which defines the classical theory that we will refer to as `loop classical cosmology'.  It is straightforward
to obtain the classical equations of motion and the pair of solutions
\begin{equation}
\nu(t)=\pm\frac{2p_\phi}{\alpha}\left[1+\frac{\alpha^4B^2}{\hbar^2}\,t^2\right]^{1/2},\label{nulcc}
\end{equation}
valid on the constraint surface $H_\mathrm{LCC}=0$, with the boundary condition that the minimum volume
occurs at $t=0$. The maximum density
\begin{equation}
\rho_\mathrm{max}=\frac{2B^2p_\phi^2}{\nu_\mathrm{min}^2}=\frac{\alpha^2B^2}{2}=\frac{\sqrt{3}}{16\pi^2\gamma^3G^2\hbar}
\end{equation}
is the same as that found in \cite{acs} for the quantum theory, while the characteristic time $\hbar/\alpha^2B$ is
of the order of the Planck time.

\subsection{Dirac observables}
`Polymer' quantization schemes of the kind used in \cite{aps2,acs} promote the Poisson bracket (\ref{sigmanupb}) to the
commutation relation
\begin{equation}
[\hat{\mathfrak{h}},\hat\nu]=-\hat{\mathfrak{h}},
\end{equation}
where $\mathfrak{h}$ is the holonomy $\mathfrak{h}=\rme^{\rmi\sigma}$, and take the operators $\hat{\mathfrak h}$
and $\hat\nu$ to act in a Hilbert space of functions $\Psi(\nu)$, which have support on a countable subset of the
real line,
\begin{equation}
\hat\nu\Psi(\nu)=\nu\Psi(\nu),\qquad\hat{\mathfrak h}\Psi(\nu)=\Psi(\nu+1).
\end{equation}
The gravity + matter system has the Hamiltonian
\begin{equation}
\hat{H}_0=\frac{\alpha^2B}{16}\hat{C}+B\vert\hat\nu\vert^{-1}\hat{p}_\phi^2,\label{H0lqc}
\end{equation}
where, with a suitable ordering of non-commuting operators in the quantized version of (\ref{Hgrav}), the action
of $\hat{C}$ is
\begin{equation}
\hat{C}\Psi(\nu)=\vert\nu+2\vert\Psi(\nu+4)+\vert\nu-2\vert\Psi(\nu-4)-(\vert\nu+2\vert+\vert\nu-2\vert)\Psi(\nu).
\label{ConPsi}
\end{equation}

As before, we want to add to this system a clock, with Hamiltonian $\hat{h}$ which, along with its conjugate
variable $\hat{t}_0$, commutes with $\hat\nu$ and $\hat{\mathfrak h}$.  Again, the total constraint is
$\hat{H}_0+\hat{h}=0$, and we would like to introduce time-dependent families of
Dirac observables (\ref{diracF}) that commute with this constraint. However, a direct construction of these
observables similar to that described in section \ref{DO} is likely to be feasible only when their time
dependence is polynomial (or perhaps has some other simple form). Here we adopt the following, somewhat less
direct strategy.

We consider an auxiliary Hilbert space $\mathcal{H}_\mathrm{aux}=\mathcal{H}_0\otimes\mathcal{H}_\mathrm{clock}$
where the Hilbert space associated with the clock is, say, $\mathcal{H}_\mathrm{clock}=L^2(\mathbb{R},\rmd h)$.
The Hilbert space $\mathcal{H}_0$ is a space of functions $\Psi(\nu,p_\phi)$, on which we again take $\hat{p}_\phi$
to act by multiplication, and on which we need the Hamiltonian $\hat{H}_0$ to be self-adjoint. This space can
itself be decomposed as $\mathcal{H}_0=\mathcal{H}_\mathrm{grav}\otimes\mathcal{H}_\phi$, with
$\mathcal{H}_\phi=L^2(\mathbb{R},\rmd p_\phi)$.  In order for $\hat{H}_0$ to be self-adjoint on $\mathcal{H}_0$,
we need both $\vert\hat\nu\vert$ and $\hat{C}$ to be self-adjoint on $\mathcal{H}_\mathrm{grav}$. They must, in 
particular, be symmetric, and this requires the inner product
$$
(\Psi_1,\Psi_2)_\mathrm{grav}=\sum_\nu \bar{\Psi}_1(\nu)\Psi_2(\nu),
$$
but again we will not attempt to establish their self-adjointness.
This inner product is \textit{different} from the one used in \cite{aps2,acs}, namely $(\Psi_1,\Psi_2)=
\sum_\nu \vert\nu\vert\bar{\Psi}_1(\nu)\Psi_2(\nu)$, which is needed for self-adjointness of the operator
$\hat\Theta=-\hat{C}\vert\hat\nu\vert$ whose positive square root is the generator of displacements
in the scalar field $\phi$, regarded as an internal `time'.

For any given value of $p_\phi$, the Hamiltonian $\hat{H}_0'(p_\phi)$, obtained by replacing $\hat{p}_\phi$ in
(\ref{H0lqc}) with its eigenvalue $p_\phi$ is a symmetric operator on $\mathcal{H}_\mathrm{grav}$.
We assume (without attempting a rigorous proof) that it has a set of $\delta$-normalized eigenfunctions $\Phi_E(\nu,p_\phi)$
\begin{equation}
\hat{H}_0'(p_\phi)\Phi_E(\nu,p_\phi)=E\Phi_E(\nu,p_\phi),\qquad(\Phi_E,\Phi_{E'})_\mathrm{grav}=\delta(E-E'),
\label{eigenvalueeqn}
\end{equation}
and that any function in $\mathcal{H}_\mathrm{aux}$ can be expressed as
\begin{equation}
\Psi(\nu,p_\phi,h)=\int\rmd E\Phi_E(\nu,p_\phi)\psi_E(h,p_\phi),\label{Psinuh}
\end{equation}
with
\begin{equation}
\psi_E(h,p_\phi)=\left(\Phi_E(p_\phi),\Psi(p_\phi,h)\right)_\mathrm{grav}.
\end{equation}
Suppose that the action of an operator $\hat{f}$ on $\mathcal{H}_0$ (equivalently, an operator on $\mathcal{H}_\mathrm{aux}$
that acts as the identity on $\mathcal{H}_\mathrm{clock}$) can be specified by a kernel $f(E,p_\phi;E',p_\phi')$.
That is,
\begin{equation}
\hat{f}\Psi(\nu,p_\phi,h)=\int\rmd E\Phi_E(\nu,p_\phi)\psi_E^{(f)}(h,p_\phi),
\end{equation}
with
\begin{equation}
\psi^{(f)}_E(h,p_\phi)=\int\rmd E'\int\rmd p_\phi'f(E,p_\phi;E',p_\phi')\psi_{E'}(h,p_\phi').
\end{equation}
The action of the corresponding Dirac observable $\hat{F}(\tau)$ defined in (\ref{diracF}) is easily found to be
\begin{equation}
\psi^{(F(\tau))}_E(h,p_\phi)=\int\rmd E'\int\rmd p_\phi'\rme^{\rmi(E-E')\tau/\hbar}f(E,p_\phi;E',p_\phi')\psi_{E'}(h+E-E',p_\phi').
\label{Fonpsi}
\end{equation}
If $\Psi(\nu,p_\phi,h)$ is a solution of the constraint equation, then the expansion coefficient in
(\ref{Psinuh}) has the form
\begin{equation}
\psi_E(h,p_\phi)=\psi(E,p_\phi)\delta(E+h),
\end{equation}
the function $\psi(E,p_\phi)$ specifying a particular solution. Because $\hat{F}(\tau)$ \textit{is} a Dirac observable,
its action (\ref{Fonpsi}) on one solution of the constraint equation yields another solution, specified by
the function
\begin{equation}
\psi^{(F(\tau))}(E,p_\phi)=\int\rmd E'\int\rmd p_\phi'f(E,p_\phi;E',p_\phi')\rme^{\rmi(E-E')\tau/\hbar}\psi(E',p_\phi').
\label{FonHphys}
\end{equation}
Evidently, the physical Hilbert space $\mathcal{H}_\mathrm{phys}$ is a space of functions $\psi(E,p_\phi)$, equipped
with an inner product chosen so as to confer self-adjointness on some class of operators, and the action of the
Dirac observables on $\mathcal{H}_\mathrm{phys}$ is specified by (\ref{FonHphys}). If the kernel can be expressed
in terms of a differential operator $\mathcal{F}$ as
\begin{equation}
f(E,p_\phi;E',p_\phi')=\mathcal{F}(E',p_\phi',\partial_{E'},\partial_{p_\phi'})\delta(E-E')\delta(p_\phi-p_\phi'),
\end{equation}
then $\psi^{(F(\tau))}(E,p_\phi)=\rme^{\rmi E\tau/\hbar}\mathcal{F}\rme^{-\rmi E\tau/\hbar}\psi(E,p_\phi)$, and we recover an
algebra of differential operators analogous to (\ref{vhath})-(\ref{H0h}), with $h=-E$.

If $\hat{f}$ is constructed from $\hat{\nu}$, $\hat{\mathfrak{h}}$ and $\hat{p}_\phi$, but does not contain $\hat\phi$,
it can be construed as an operator on $\mathcal{H}_\mathrm{grav}$ that depends parametrically on $p_\phi$. Its
kernel has the form $f(E,p_\phi;E',p_\phi')=f(E,E',p_\phi)\delta(p_\phi-p_\phi')$, with
\begin{equation}
f(E,E',p_\phi)=(\Phi_E(p_\phi),\hat{f}\Phi_{E'}(p_\phi))_\mathrm{grav},\label{simplekernel}
\end{equation}
and we will use this expression to estimate the volume operator $\hat{V}(\tau)$.
\subsection{Eigenfunctions of $\hat{H}_0'(p_\phi)$}
As described in \cite{acs,acsrev}, the constraint equation $\hat{H}_0\Psi=0$ that applies in the absence of the
clock can be solved exactly if we restrict $\mathcal{H}_\mathrm{grav}$ to functions $\Psi(\nu)$ that have support
only at $\nu=4n$, $n\in\mathbb{Z}$; the operator $\hat{C}$ in (\ref{ConPsi}) clearly has a well-defined action on
this restricted space.  We will also restrict attention to this sector, but are able to obtain only approximate
solutions to the eigenvalue equation (\ref{eigenvalueeqn}) for nonzero eigenvalues $E$.

With the definition
\begin{equation}
\Phi_E(\nu,p_\phi)=\frac{\vert\nu\vert}{4}\tilde\chi_E(n,p_\phi)=\vert n\vert\tilde\chi_E(n,p_\phi)
\end{equation}
the eigenvalue equation we wish to solve is
\begin{equation}
\frac{\alpha^2B}{8}\left[(n+1)(2n+1)\tilde\chi_E(n+1)+(n-1)(2n-1)\tilde\chi_E(n-1)-4n^2\tilde\chi_E(n)\right]
+\frac{Bp_\phi^2}{4}\tilde\chi_E(n)=E\vert n\vert\tilde\chi_E(n),
\end{equation}
where we suppress the argument $p_\phi$ of $\tilde\chi_E$, and on taking the Fourier transform
\begin{equation}
\tilde\chi_E(n)=\frac{1}{\pi}\int_0^\pi\rmd k\rme^{-\rmi2nk}\chi_E(k),\qquad\chi_E(k)=\sum_{n=-\infty}^\infty
\rme^{\rmi2nk}\tilde\chi_E(n)
\end{equation}
this becomes
\begin{equation}
B\left(\alpha^2\sin k\partial_k\sin k\partial_k+p_\phi^2\right)\chi_E(k)=E\left\vert-2\rmi\partial_k\right\vert\chi_E(k).
\end{equation}
Finally, the change of variable $\sin k=1/\cosh x$ yields (with an obvious economy of notation)
\begin{equation}
(\partial_x^2+\lambda^{-2})\chi_E(x)=\tilde{E}\left\vert-2\rmi\cosh x\partial_x\right\vert\chi_E(x),\label{eveqn}
\end{equation}
where we have defined $\lambda=\alpha/p_\phi$ and $\tilde{E}=E/\alpha^2B$.  With this parametrization, the inner
product in (\ref{eigenvalueeqn}) is
\begin{equation}
(\Phi_E,\Phi_{E'})_\mathrm{grav}=
\frac{1}{4\pi}\int_{-\infty}^\infty\rmd x\,\cosh x\partial_x\bar\chi_E(x)\partial_x\chi_{E'}(x).
\end{equation}

In the form (\ref{eveqn}), the eigenvalue equation is trivial when $E=0$, but hard to solve when $E$ is nonzero.
We will obtain an approximate solution by means of an expansion in powers of $\lambda$, but our analysis will
be largely devoid of rigour.  In practical terms, the approximation is likely to be quite good; for example,
we have $\lambda\sim 10^{-123}$ for the state considered in section \ref{viab1}.  Consider the ansatz
\begin{equation}
\chi_E(x)=-\rmi\lambda\mathcal{N}\exp\left[\rmi\lambda^{-1}x +\rmi f_E(x)+\lambda g_E(x)\right],\label{ansatz}
\end{equation}
where $f_E(x)$ and $g_E(x)$ are real functions expressible as power series in $\lambda$,
\begin{equation}
f_E(x)=f_0(x)+\lambda f_1(x)+\ldots,\qquad g_E(x)=g_0(x)+\lambda g_1(x)+\ldots.
\end{equation}
The prefactor $-\rmi\lambda$ is inserted for later convenience and $\mathcal{N}$ is a normalization constant.
We wish, of course, to determine the functions $f_E(x)$ and $g_E(x)$ by substituting this ansatz into (\ref{eveqn}),
but the absolute value of the volume operator $\hat\nu=-2\rmi\cosh x\partial_x$ presents a difficulty.  It is not
hard to show that this operator is positive on the space of functions whose Fourier transforms $\int\rmd x \rme^{-\rmi\omega x}
\chi(x)$ have support only for $\omega>0$ and \textit{vice versa}, and that these two spaces are orthogonal.
Clearly, the function $\rme^{\rmi\lambda^{-1}x}$ lies in the positive-volume space, and one might hope that the
same is true of the trial function (\ref{ansatz}), at least within the expansion in powers of $\lambda$, but we
are not able to prove this. We proceed by removing the absolute value symbol in (\ref{eveqn}), and verifying
\textit{a posteriori} the positivity of $\hat\nu$ on the space spanned by the resulting set of approximate
eigenfunctions.

With this simplification, we find
\begin{eqnarray}
f_E(x)&=&-\tilde{E}\sinh x+\frac{\lambda}{4}\tilde{E}^2(\cosh x \sinh x +x)+\mathrm{O}(\lambda^2)\\
g_E(x)&=&\frac{1}{2}\tilde{E}\cosh x -\frac{\lambda}{4}\tilde{E}^2\cosh^2x+\mathrm{O}(\lambda^2).
\end{eqnarray}
To make systematic use of this expansion, we construct the polar representation of the function
\begin{equation}
\xi_E(x):=\partial_x\chi_E(x)=\rho_E(x)\rme^{\rmi\theta_E(x)},
\end{equation}
with the result
\begin{eqnarray}
\rho_E(x)&=&\mathcal{N}\left[1-\frac{\lambda}{2}\tilde{E}\cosh x+\mathrm{O}(\lambda^2)\right]\label{rhoseries}\\
\theta_E(x)&=&\lambda^{-1}x-\tilde{E}\sinh x+\frac{\lambda}{4}\tilde{E}^2(\cosh x\sinh x +x)+\mathrm{O}(\lambda^2).
\label{thetaseries}
\end{eqnarray}
An important check on the consistency of our procedure is now to obtain the inner product
\begin{equation}
(\Phi_E,\Phi_{E'})_\mathrm{grav}=\frac{1}{4\pi}\int_{-\infty}^\infty\rmd x\,\cosh x\,\bar{\xi}_E(x)\xi_{E'}(x)
=\frac{1}{4\pi}\int_{-\infty}^\infty\rmd x\,\cosh x\,\rho_E(x)\rho_{E'}(x)\rme^{\rmi[\theta_E(x)-\theta_{E'}(x)]}.
\label{deltaint}
\end{equation}
Self-adjointness of $\hat{H}_0$ requires that this be proportional to $\delta(E-E')$, which in turn requires
\begin{equation}
(E-E')^{-1}\partial_x\left[\theta_E(x)-\theta_{E'}(x)\right]\propto\cosh x\,\rho_E(x)\rho_{E'}(x),
\end{equation}
in order that the final integral in (\ref{deltaint}) reduce to $\int\rmd s\rme^{\rmi(E-E')s}$. This can be
checked order by order in $\lambda$, and is readily seen to hold at the order of our explicit calculations.
We find that $(\Phi_E,\Phi_{E'})_\mathrm{grav}=\delta(E-E')$, if the normalization constant in (\ref{ansatz})
is $\mathcal{N}=(2/\alpha^2B)^{1/2}$.

Matrix elements of the volume operator are given by
\begin{equation}
(\Phi_E,\hat\nu\Phi_{E'})_\mathrm{grav}=\frac{1}{4\pi}\int_{-\infty}^\infty\rmd x\,\cosh x\,\bar{\xi}_E(x)\xi_{E'}^{(\nu)}(x)
\end{equation}
with
\begin{equation}
\xi_E^{(\nu)}(x)=\partial_x\left[\hat\nu\chi_E(x)\right]=-2\rmi\partial_x\left[\cosh x\,\xi_E(x)\right].
\end{equation}
To lowest order, we find
\begin{equation}
\xi_E^{(\nu)}(x)
= 2\lambda^{-1}\cosh x\xi_E(x)=2\lambda^{-1}\left(1-\alpha^4B^2\partial_E^2\right)^{1/2}\xi_E(x)\label{nume}
\end{equation}
and thus
\begin{equation}
(\Phi_E,\hat\nu\Phi_{E'})_\mathrm{grav}=2\lambda^{-1}\left(1-\alpha^4B^2\partial_{E'}^2\right)^{1/2}\delta(E-E').
\end{equation}
It follows, at this order of approximation, that $\hat\nu$ is positive, $(\Psi,\hat\nu\Psi)_\mathrm{grav}>0$ on the
space spanned by these eigenfunctions; obviously, by starting with $\chi_E(x)=\exp(-\rmi\lambda^{-1}x+\ldots)$ we would
find a complementary space on which $\hat\nu$ is negative.  The form of the series (\ref{rhoseries}) and (\ref{thetaseries})
suggests that a similar approximation might be obtained by expanding in the eigenvalue $E$, which we also want to be
small, since it is minus the energy of a test clock.  The leading terms of such an expansion do agree with those
obtained above, but we have not been able to develop it in a wholly consistent manner.
\subsection{Time evolution of the volume\label{tev}}
The matrix element (\ref{nume}) together with (\ref{FonHphys})-(\ref{simplekernel}) yields the expectation value
of the Dirac observable associated with $\hat\nu$ as
\begin{equation}
\langle\hat\nu(\tau)\rangle=\int\rmd E\int\rmd E'\int\rmd p_\phi\,\bar\psi(E,p_\phi)\psi(E',p_\phi)
\int_{-\infty}^\infty\frac{\rmd s}{2\pi\hbar}\rme^{-\rmi(E-E')s/\hbar}\frac{2p_\phi}{\alpha}\left(1+\frac{\alpha^4B^2}{\hbar^2}(s+\tau)^2
\right)^{1/2},
\end{equation}
where we observe an obvious correspondence with the classical solution (\ref{nulcc}).

We would again like to assess the viability of a test clock, by restricting the clock's energy $-E$ to a small
range of values around a mean value $\epsilon_\mathrm{c}$, which is itself small. As in section \ref{viab1},
we take the dispersion in $p_\phi$ to be negligible and use the Gaussian wavefunction
\begin{equation}
\psi(E)=\left(\frac{\tilde\Delta}{\pi}\right)^{1/4}\rme^{-\tilde\Delta(E+\epsilon_\mathrm{c})^2/2}.\label{wfn}
\end{equation}
This state differs in detail from (\ref{psiofeta}); roughly speaking, the two variances are related by
$\tilde\Delta=\Delta/\alpha^2\epsilon^2_\mathrm{c}$. Taking into account the rescaling $v=(2B)^{-1}\vert\nu\vert$, we
find for the Dirac observable $\hat{V}(\tau)$ associated with the volume $\hat{v}\sim\widehat{a^3}$
\begin{equation}
V(\tau):=\langle\hat{V}(\tau)\rangle=v_0\frac{1}{\sqrt{\pi}}\int_{-\infty}^\infty\rmd s\,\rme^{-s^2}
\left[1+\tau_\mathrm{P}^{-2}\left(\tau+\sqrt{\tilde\Delta}\hbar s\right)^2\right]^{1/2}
\end{equation}
where $v_0=p_\phi/\alpha B$ is the minimum (`bounce') volume attained in the classical solution, and
$\tau_\mathrm{P}=\hbar/\alpha^2B\approx 6.8\times 10^{-45}\mathrm{s}$ is of the order of the Planck time.
In this state, the bounce occurs at $\tau=0$, but including a factor $\rme^{-\rmi\tau_0E/\hbar}$ in
the wavefunction (\ref{wfn}) would displace it to $\tau=-\tau_0$. We also need
\begin{equation}
\langle\hat{V}(\tau)^2\rangle=v_0^2\frac{1}{\sqrt{\pi}}\int_{-\infty}^\infty\rmd s\,\rme^{-s^2}
\left[1+\tau_\mathrm{P}^{-2}\left(\tau+\sqrt{\tilde\Delta}\hbar s\right)^2\right]=v_0^2\left(1+\frac{\tilde{\Delta}
\hbar^2}{2\tau_\mathrm{P}^2}+\frac{\tau^2}{\tau_\mathrm{P}^2}\right).
\end{equation}

If the clock's energy is to be restricted to a small range near $\epsilon_\mathrm{c}$, we need
$\tilde{\Delta}\gg\epsilon_\mathrm{c}^{-2}$, or $\Delta\gg\alpha^2$, which reproduces the first
inequality in (\ref{tccomp}).

Now consider a late time, $\tau\gg\tau_\mathrm{P}$, at which it ought to be
possible to set up a quasi-classical state. If $\tilde{\Delta}\hbar^2/\tau_\mathrm{P}^2$ is not too
large (and we shall soon see that this quantity should be small), the volume can be approximated as
\begin{equation}
V(\tau)=\frac{v_0\tau}{\tau_\mathrm{P}}\left[1+\frac{1}{2}\frac{\tau_\mathrm{P}^2}{\tau^2}
+O\left(\frac{\tau_\mathrm{P}^4}{\tau^4}\right)\right].\label{vlate}
\end{equation}
This reproduces the second term of (\ref{exvoftau}), linear in $\tau$, but misses the last term proportional
to $\tau^2$ which, as discussed earlier, is important at very late times when the clock energy dominates that
of the scalar field.  One would expect the large-volume evolution to be insensitive to the
quantization scheme, and it is likely that the discrepancy indicates a failure of the approximations
(\ref{rhoseries}) and (\ref{thetaseries}) for large values of $E$. With the approximation (\ref{vlate}),
we estimate the dispersion in the volume at late times as
\begin{equation}
\left(\frac{\delta V}{V}\right)^2\approx \frac{v_0^2\tilde{\Delta}\hbar^2}{2V^2\tau_\mathrm{P}^2}
\approx\frac{\Delta E_\phi}{2\epsilon_\mathrm{c}^2V},
\end{equation}
and requiring this to be small reproduces the second inequality in (\ref{tccomp}).  At late times, therefore,
the criterion for the notion of a `test clock' to be viable is the same in the LQC and Wheeler-de-Witt
quantization schemes, as it should be.

At the `bounce' volume, which occurs at $\tau=0$, we might expect this criterion to be modified. Assuming
that $\tilde{\Delta}\hbar^2/\tau_\mathrm{P}^2$ is small, we obtain the approximation
\begin{equation}
V(0)=v_0\left[1+\frac{1}{4}\left(\frac{\tilde{\Delta}\hbar^2}{\tau_\mathrm{P}^2}\right)
-\frac{3}{32}\left(\frac{\tilde{\Delta}\hbar^2}{\tau_\mathrm{P}^2}\right)^2+\ldots\right]
\end{equation}
and
\begin{equation}
\left(\frac{\delta V}{V}\right)^2\approx\frac{1}{8}\left(\frac{\tilde{\Delta}\hbar^2}{\tau_\mathrm{P}^2}\right)^2.
\end{equation}
In order to have this dispersion small, and to have the clock's energy close to $\epsilon_\mathrm{c}$, we
need
\begin{equation}
\epsilon_\mathrm{c}^{-2}\ll\tilde{\Delta}\ll\epsilon_\mathrm{p}^{-2},
\end{equation}
where $\epsilon_\mathrm{P}=\hbar/\tau_\mathrm{P}\approx 1.6\times10^{10}\mathrm{J}$ is about 10
times the Planck energy.  Evidently, independently of the quantum state, the clock's energy
$\epsilon_\mathrm{c}$ must be much larger than $\epsilon_\mathrm{P}$. This condition might be
regarded as satisfied by the 1kg clock considered in section \ref{viab1}, but not by a huge margin. Whether
$\epsilon_\mathrm{c}$ is much smaller than the scalar-field energy depends, of course on the particular
state considered.  For the quasi-classical state of a region of size $10^{80}\mathrm{m}^3$ considered in
section \ref{viab1}, the constant $p_\phi^2=2V^2\rho_\phi$ is about $10^{151}\mathrm{J\,m}^3$.  At its
`bounce' volume, $v_0$, this region contains a scalar field energy $E_\phi=p_\phi^2/2v_0
=\epsilon_\mathrm{P}p_\phi/2\alpha\approx10^{122}\epsilon_\mathrm{P}$, so the condition
$\epsilon_\mathrm{P}\ll\epsilon_\mathrm{c}\ll E_\phi$ is not hard to satisfy.
\section{Discussion\label{discussion}}
In section \ref{reltime}, we identified two features of a widely adopted relational approach to time
evolution in generally-covariant systems, which we believe to be shortcomings of this approach. In the
context of simplified time-reparametrization-invariant models such as the cosmological model studied
in this paper, evolution with respect to an `internal time' is described by operators of the form
(\ref{vofphi0}).  These operators depend parametrically on some variable (in this case $\phi_0$) associated
with a dynamical variable (in this case a scalar field) which is regarded as a physical clock. The
shortcomings (in our view) of this account are two-fold: (i) it provides no `heraclitian' time that
would serve to make sense of, for example, the everyday experience that a well-constructed clock
reads `10s' about ten seconds after it read `0'; (ii) a parameter such as $\phi_0$ cannot be interpreted
as an observed value of the chosen physical clock variable, because no dynamical quantity exists that
might be observed to have this value, either as a function on the reduced classical phase space, or
as an operator on the physical Hilbert space of the quantized theory.

In sections \ref{conqu} and \ref{loopqu}, we described, within two standard quantization schemes,
a possible means of alleviating these difficulties by introducing a `test clock' that is to be
regarded as internal to some specific observer (or observing apparatus). In this way, we could construct
classical Dirac observables, or `evolving constants of the motion' (\ref{classDiracF}) and their
quantum-mechanical counterparts (\ref{diracF}).  These gauge-invariant quantities evolve with
respect to a heraclitian time $\tau$, their evolution being governed \textit{exactly} by the usual
Hamilton or Heisenberg equations of motion. Like the physical clocks in other relational schemes,
the test clock is not represented by any independent Dirac observable. Intuitively, this makes sense,
insofar as the clock is internal to the observer from whose point of view the time evolution is
described, and thus inaccessible to that observer.  Problem (ii) above does not arise, because $\tau$
has the textbook interpretation of the geometrical proper time that elapses along this observer's
worldline, and does \textit{not} correspond to any reading of the clock.

It is worth emphasizing that more is at stake than interpretation.  In constructing the physical Hilbert
space, one naturally wants time evolution to be generated by a self-adjoint operator.  As noted
earlier, different inner products, and to that extent different quantum theories, are needed to confer
self-adjointness on the generator of evolution in the scalar field $\phi$ or on the original
Hamiltonian $H_0$, which generates evolution in $\tau$.  It seems to us that self-adjointness of $H_0$
is a more natural requirement. Any specific choice of an internal time variable is essentially an
arbitrary matter, and it seems unnatural that the resulting quantum theory should depend on that
arbitrary choice.

If the notion of a `test clock'
is to make good sense, it should be possible to restrict the clock's energy to be a negligible fraction
of the total.  This places limits on the resolution with which the values of time-dependent observables
might be determined, but we found in sections \ref{viab1} and \ref{tev} that reasonable compromises
are possible. That is, one can take the clock's energy to be reasonably small, while leaving the
time-dependent observables reasonably sharply defined.

As noted in section \ref{hctc}, our construction of time-dependent Dirac observables is similar,
but not identical, to relational constructions adopted by several previous authors, and it is
worth comparing the two approaches in some detail.  For classical systems, the original idea of
Rovelli \cite{rovtime,rovref,rovpartial} has been developed in considerable generality by
Dittrich \cite{dittrich06,dittrich07} (see \cite{pons97,pons05,pons09} for a somewhat different
perspective), and the quantum theory obtained by reduced-phase-space quantization is discussed
in \cite{thiemann06,giesel10}. In models of the kind discussed here, these constructions amount to
replacing the function $t_0(\bm{x})$ defined by (\ref{t0def}) with a function $t_1(\bm{x},\rho)$
defined by
\begin{equation}
r(\bar{\bm{x}}(\bm{x},t_1))=\rho,\label{t1def}
\end{equation}
and the Dirac observable $F(\tau)$ defined in (\ref{Foftau}) with
\begin{equation}
F_1(\rho)=\bar{f}(\bm{w},t_1(\bm{x},\rho)).\label{F1def}
\end{equation}
It is again straightforward to show that $\{t_1,h\}=-1$ and that, for each value of the parameter
$\rho$, $F_1(\rho)$ Poisson-commutes with the total constraint $H_0+h$. Classically, \textit{before
the constraint is imposed}, $F_1(\rho)$ seems to be interpretable as the value of $f$ when the clock
reading $r(\bm{x})$ has the value $\rho$. The operator (\ref{vofphi0}) is a quantum implementation of
this Dirac observable for the case that $f$ is the volume and $r(\phi)=\phi$ the scalar field. Clearly,
this is a different construction from the one we have advocated, and we have argued that it does not
have a satisfactory interpretation: before imposing the constraint, $\rho$ seems to be an observed value
of the clock reading $r$, but no corresponding observable $R$ exists in the final, constrained theory.

There is,
however, a special case in which the net results (\ref{Foftau}) and (\ref{F1def}) of these two different
constructions are algebraically indistinguishable;
namely, when the solution of the equation of motion for $r$ is linear in $t$, so that
$t_1(\bm{x},\rho)=-r(\bm{x})+\rho$. This is true for a special kind of clock, whose Hamiltonian $h=p_r$ is
just the momentum conjugate to $r$. In that case, the total constraint has the `deparametrized' form
$H=H_0+p_r$, and the replacement $p_r=-\rmi\hbar\partial_\rho$ yields a Schr\"odinger-like equation.
It can be argued (see, for example, \cite{hartle,uw}) that the Hamiltonian $h=p_r$, which is unbounded
below, cannot describe any physical clock.\footnote{This objection is irrelevant to our construction,
because our clock is not required to provide a linear measure of proper time. As implemented in
section \ref{conqu}, the clock has a positive energy (see (\ref{h2eta})) and in section \ref{loopqu}
its energy was allowed to be negative, with small probability, purely as a matter of technical
convenience. The objection \cite{uw} that measured readings of a quantum-mechanical clock with energy bounded
below are not guaranteed to increase monotonically with time also does not apply, because our $\tau$ does
not represent any such measured values.} However, Brown and Kucha\v{r} \cite{brown} have modeled a
dust-filled universe by using a collection of scalar fields, of which one, say $T$, \textit{when the
classical equations of motion are satisfied}, is linear in the proper time along particle worldlines.
This model has been studied more recently by several authors from the point of view of the relational formalism
\cite{giesel10,giesel,giesel2,amemiya} (see also \cite{montani,montani09}).

In particular, Amemiya and Koike \cite{amemiya} investigate the quantum dynamics of a homogeneous
universe whose matter content comprises Brown-Kucha\v{r} dust together with classical radiation
and a cosmological constant.  Their Hamiltonian constraint has the form $H_0(a,p_a)+p_T$, where $p_T$,
the momentum conjugate to $T$, is essentially the total energy of the dust content of a compact
universe. Quantization amounts to replacing $p_T$ with $-\rmi\hbar\partial_T$ and $p_a$ with
$-\rmi\hbar\partial_a$ to obtain the Schr\"odinger-like equation
$\rmi\hbar\partial_T\Psi(T,a)=H_0(a,-\rmi\hbar\partial_a)\Psi(T,a)$.  More formally, the relational
formalism yields a family of Dirac observables $A(\rho)$ through the prescription (\ref{F1def}), with
$t_1(T,\rho)=\rho-T$.  Reduced-phase-space quantization then promotes these to `Heisenberg-picture' operators
$\hat{A}(\rho)$ on the physical Hilbert space and the corresponding Schr\"odinger-picture wavefunction
$\Psi(\rho,A)$ is governed by
\begin{equation}
\rmi\hbar\partial_\rho\Psi(\rho,A)=H_0(A,-\rmi\hbar\partial_A)\Psi(\rho,A).\label{amschrod}
\end{equation}
We retain the symbol $\rho$ here to emphasize that this variable is a value assigned to the scalar field
$T$, whose classical equation of motion \textit{happens to have the solution} $\bar{T}(t)=\bar{T}(0)+t$,
whereas our variable $\tau$ \textit{is}, up to a choice of origin, the arc length $t$.  Again, there is no
operator on the physical Hilbert space to represent the dust, so although $\rho$ is a value assigned
to $T$, it cannot be regarded as the result of a measurement of $T$, though in this special case it can be
loosely associated with proper time by appealing to the classical equation of motion. Of course, the
constraint implies that the energy of the dust is given by $-H_0$, but this ought not to be the whole story:
cosmologists expect to verify the Friedmann equation through \textit{independent} measurements of the Hubble
parameter and the distribution of matter.  This is a matter of some practical importance if the
Brown-Kucha\v{r} fields are used to model the actual non-relativistic matter in our universe.  In
\cite{amemiya}, the `Schr\"odinger' equation (\ref{amschrod}) is solved for conditions corresponding to an
energy density of radiation much larger than is observed, and yet their universe remains substantially
unaffected by this radiation at the earliest times, when classical cosmology would lead one to expect a
radiation-dominated universe.  We suspect that this reflects a rather large value of $\langle -H_0\rangle$
corresponding to a large density of dust, whose presence can be inferred only from the behaviour of the scale
factor. Presumably, one could arrive at a closer approximation to our universe by considering a state in
which $\langle -H_0\rangle$ is small enough, but that would not solve the interpretational problem that the
dust is not directly observable.

Alternatively, one can try to regard the Brown-Kucha\v{r} `dust' as being truly unobservable.  This is the
view taken in \cite{giesel}, and in \cite{giesel2} these authors speculate that this dust might be a candidate
for dark matter, since it interacts only gravitationally.\footnote{In fact, it is claimed in these papers
that the dust must have a negative energy density, in order  to yield a physical Hamiltonian that is
positive, but that this does not matter, because the dust is unobservable. Amemiya and Koike \cite{amemiya}
accept the negative Hamiltonian implied by a positive energy density for their dust, on the grounds that the
original gravitational Hamiltonian is in any case negative, and we think they are right to do so, at least
in the context of their homogeneous cosmology.} On this view, one has to postulate that the
universe contains, in addition to ordinary detectable matter, another species which is unobservable in
principle, not merely because it does not interact, but because it does not feature in the physical
phase space.  This seems implausible to us, because the dust appears in the first instance on the same
footing as any other matter, and it is only as a matter of technical convenience that one chooses to
solve the constraints by eliminating the variables associated with this, rather than some other species.

By contrast, the view proposed in this paper is that a specific observer, say $O_1$, will account for
time-dependent observations made along her own worldline by describing the physical phase space in terms
of variables that do not include a clock internal to herself.  It does seem plausible to us that she
cannot, in principle, make measurements of the reading of this particular clock.  If the model is
sufficiently detailed, her description of the physical phase space may include the variables corresponding
to a clock that is internal to a second observer, $O_2$, whose readings she might, in principle, be able to
measure (if a socially acceptable procedure could be agreed on).  To account for time-dependent observations
made along \textit{his} worldline, however, $O_2$ will coordinatize the physical phase space using variables
that do not include his own internal clock, though they would include the clock internal to
$O_1$.\footnote{In \cite{giesel2}, the dust is described as a `test observer medium'.  Taken literally,
this would imply a space-filling congruence of observers, who are unobservable not only to themselves
but to each other.}

We have, of course, substantiated this view of time evolution only in the context of
a greatly simplified cosmological model.  Within this model, the time parameter $\tau$, which has an
unambiguous classical interpretation as the proper time that elapses along an observer's worldline, survives
quantization as a gauge-invariant c-number parameter, because the lapse function $N(s)$ in (\ref{propertime})
is not a dynamical variable, and is not promoted to an operator in the quantum theory.\footnote{A proposal for
constructing a spacetime metric tensor in which the time coordinate is replaced with a dynamical
clock variable is described in \cite{kaminski}.  This yields a lapse function with respect to clock
time that \textit{is} a dynamical variable, but it is quite distinct from the one used here.}
However, by considering only a comoving observer, we have avoided dealing explicitly with the dynamical
variables $x^\mu(\lambda)$ in (\ref{clockaction}), which more generally are needed to specify precisely
what is meant, in the final quantum theory, by `an observer's worldline'.  To deal adequately with a
genuinely localized observer in a more general spacetime is a much more challenging enterprise, which we
plan to explore in future work.

\acknowledgements
It is a pleasure to thank Abhay Ashtekar, Bill Unruh, Gabor Kunstatter, Don Salisbury and especially Jorma Louko for
comments on earlier versions of this work, and a referee for drawing our attention to Ref.\cite{blyth}.

\end{document}